\documentclass[article,amsmath,amssymb,twocolumn,superscriptaddress]{revtex4}
\usepackage{chngcntr}
\usepackage{grffile}
\usepackage{MnSymbol}
\usepackage{textcomp}
\usepackage[normalem]{ulem}
\usepackage[dvipsnames]{xcolor}
\usepackage[colorlinks,
            linkcolor=blue,
            anchorcolor=blue,
            citecolor=blue
            ]{hyperref}

\begin{document}

\title{Eigenstate plateau transition and equilibration in\\ one-dimensional quantum lattice models}
\author{Wei-Han Li}\email{wei-han.li@outlook.com}
\affiliation{Max Planck Institute for the Physics of Complex Systems, 01187 Dresden, Germany}
\author{Abbas Ali Saberi}\email{ab.saberi@ut.ac.ir,\;saberi@pks.mpg.de}
\affiliation{Department of Physics, University of Tehran, P. O. Box 14395-547, Tehran, Iran}
\affiliation{Max Planck Institute for the Physics of Complex Systems, 01187 Dresden, Germany}

\begin{abstract}
We report a remarkable spectral phenomenon in a generic quantum lattice gas model. As the interaction strength increases, eigenstates spontaneously reorganize, leading to plateaus in the interaction energy. Gaps between plateaus can emerge similarly to continuous phase transitions. Our perturbation analysis indicates that, aside from these gaps, the plateau structure of the spectrum remains stable in the thermodynamic limit. The structured spectrum naturally manifests in far-from-equilibrium dynamics, exhibiting multiple stages. Our findings reveal a connection between emergent properties in high-energy states and equilibration dynamics in closed quantum systems, offering insights into the impact of interactions across the entire energy spectrum. These results directly relate to experiments probing equilibration in quantum spin systems and lattice gases.
\end{abstract}
\maketitle


\section{Introduction} 
The study of quantum many-body systems has mainly focused on their equilibrium properties and low-energy characteristics, particularly ground states and associated quantum phase transitions. However, exploring exotic out-of-equilibrium phenomena in the vast high-energy realms delves into fundamental questions that are still unanswered. 
Recent years have seen a burgeoning interest in exploring the equilibration dynamics of closed quantum systems \cite{Eisert2015, Gogolin2016}, particularly in low dimensions, propelled by advances in ultracold atomic experiments \cite{Langen2015, Gross2017, Bloch2008, Bloch2012}. This progress has enabled the realization of archetypal model Hamiltonians \cite{Greiner2002, Stoferle2004, Kohl2005, Spielman2007, Esslinger2010} and the ability to study dynamical equilibration phenomena such as mass expansion \cite{Scheider2012, Trotzky2012, Ronzheimer2013, Xia2015, Scherg2018} and spin transport \cite{Hild2014, Jepsen2020} in highly controllable settings. 
A key frontier has been the exploration of long-range interacting quantum gases \cite{Defenu2023}, where strong inter-site interactions can be engineered in optical lattices, giving rise to effective long-range spin models \cite{Paz2013, Patscheider2020, Scholl2021} and extended Hubbard models (EHMs) \cite{Baier2016, Su2023, Sanchez2021}. 
The delicate interplay between these tunable interactions and kinetic energy in such systems can lead to novel quantum phases \cite{Torre2006, Deng2011, Lahaye2009, Dutta2015} and rich non-equilibrium dynamical behavior \cite{Barbiero2015, WHL1, WHL2, WHL3, Korbmacher2023}.

For various one-dimensional (1D) EHMs and spin models, previous studies revealed that intersite interactions strong enough to drive ground-state phase transitions can induce intriguing changes in dynamical properties. These changes include a qualitative slowdown of quench dynamics \cite{Sanchez2021}, distinct features in the self-return probability \cite{Misguich2016}, and transitions in far-from-equilibrium transport behavior \cite{Steinigeweg2017, Ljubotina2017}. 
In this work, we consider a generic 1D EHM, which can be mapped to the XXZ model (spin-$\frac{1}{2}$ Heisenberg chain). Our main focus lies in exploring the emergent properties of the many-body eigenstates arising from the tunable intersite interactions and identifying their signatures in the out-of-equilibrium dynamics.

We uncover a hidden structure across the spectrum as intersite interactions increase, in which eigenstates exhibit plateaus of interaction energies. The gaps between these plateaus can play a role similar to order parameters in continuous phase transitions. Nonetheless, the plateau structure of the system is not contingent on the presence of the gaps. Using perturbation theory, we estimated a threshold interaction strength, beyond which the spectrum exhibits a stable structure of plateaus at the thermodynamic limit. Our analysis further predicts statistical shifts in the distribution of eigenstates, as described by extreme value theory. Although probing the steady feature of an eigenstate with specific energy may be impossible in a thermodynamic system, the full spectrum's impact influences far-from-equilibrium dynamics involving numerous eigenstates. We study the time evolution of Fock states, revealing multiple equilibration stages that universally apply across different interaction forms in current experiments. Our findings offer deeper insights into previous theoretical studies \cite{Misguich2016, Steinigeweg2017, Ljubotina2017} and a recent experimental observation \cite{Sanchez2021}.

\section{The model}
We consider the EHM described by the Hamiltonian 
\begin{equation}\label{tv_H}
H = \sum_{j} 
- t \left( \hat{a}^{\dagger}_{j+1}\hat{a}_{j} + \hat{a}^{\dagger}_{j}\hat{a}_{j+1} \right)
+ V \hat{n}_{j+1}\hat{n}_{j},
\end{equation}
where $\hat{a}^{\dagger}_j$ ($\hat{a}_j$) is the creation (annihilation) operator at lattice site $j$ and $\hat{n}_j = \hat{a}^{\dagger}_j\hat{a}_j$ is the on-site occupation with the hard-core constraint $(\hat{a}^{\dagger}_j)^2=0$. 
We assume periodic boundary conditions, and half filling, $N/L=\frac{1}{2}$, for $N$ particles moving in $L$ sites.

Equation (\ref{tv_H}) is a generic model capturing 1D quantum lattice gases that emphasize nearest-neighbor (NN) interactions. 
The interplay between the hopping amplitude $t$ and the interaction strength $|V|$ is the core factor influencing the eigenstates. 
A well-known example is the quantum phase transition \cite{Nozieres2004} when the EHM is mapped onto the XXZ model \cite{note_mapping}. When $|V|/t$ exceeds $2$, the ground state changes from paramagnetic to ferromagnetic or antiferromagnetic depending on the sign of $V$. Nonetheless, an increase in $|V|$ does not merely impact the ground state; in principle, the influence extends to all excited states in the spectrum.

\subsection*{Observables and parameters}
Our study focuses on how excited states respond to $V$, the control parameter. To capture this response, we define the parameter $\hat{l} = \frac{\partial H}{\partial V}$, where
\begin{equation}\label{l}
    \hat{l} = \sum_j \hat{n}_{j+1} \hat{n}_j
\end{equation}
represents the number of NN links. This quantity can be mapped to the number of domain walls and serves as a conserved quantum number in the limit $V \to \infty$ \cite{Sharma2014}.
The ferromagnetism (antiferromagnetism) mentioned above thus corresponds to the extreme expectation value $\langle \hat{l} \rangle\to N-1$ ($\langle \hat{l} \rangle\to 0$). 
States with the same $\langle \hat{l} \rangle$ may have different numbers of solitary particles or holes, e.g., $\bullet\bullet\bullet\circ\bullet\circ$ and $\bullet\bullet\circ\circ\bullet\bullet$. The number of these particles and holes is
\begin{equation}
\label{s}
\hat{s} = \sum_{j}(\hat{n}_{j+1}-1) \hat{n}_{j} (\hat{n}_{j-1}-1) + \hat{n}_{j+1} (\hat{n}_{j}-1) \hat{n}_{j-1}, 
\end{equation}  
where the first and second terms denote the solitary particles and holes, respectively. 
 
A solitary particle in Eq. (\ref{s}) can move freely when no links are altered, as in the case of $\circ\bullet\circ\circ \leftrightarrow \circ\circ\bullet\circ$. A similar motion occurs for a hole when one link is broken and another is created, as in $\bullet\bullet\circ\bullet \leftrightarrow \bullet\circ\bullet\bullet$. Through these motions, a solitary particle can turn into a solitary hole and vice versa, for example, $\bullet\bullet\bullet\circ\circ\bullet\circ \leftrightarrow \bullet\bullet\bullet\circ\bullet\circ\circ \leftrightarrow \bullet\bullet\circ\bullet\bullet\circ$.  
Therefore, double counting of solitary particles and holes must be avoided. Their total number is constrained by
\begin{equation}
\label{N}
    N = \hat{l} + \hat{s} + \hat{c},
\end{equation}
where $\hat{c}$ represents the number of clusters. A cluster is defined as a particle array without holes or as a particle configuration that can be transformed into such an array through moving solitary particles or holes.
 

From now on, we will refer to the solitary particles and holes together as \textit{singlons}. 
The motion of singlons conserves the number of NN links, thus in the regime $|V|>t$ an eigenstate is expected to be mainly spanned by a set of Fock bases mutually connected through singlon movement. This relationship suggests the number of NN links is the main parameter to characterize the eigenstate structures when $|V|$ increases, \textcolor{blue}{see Appendix A.}

\section{ Plateau transitions} 
\label{plateau_gaps}
\begin{figure}[t]
\includegraphics[width=\columnwidth]{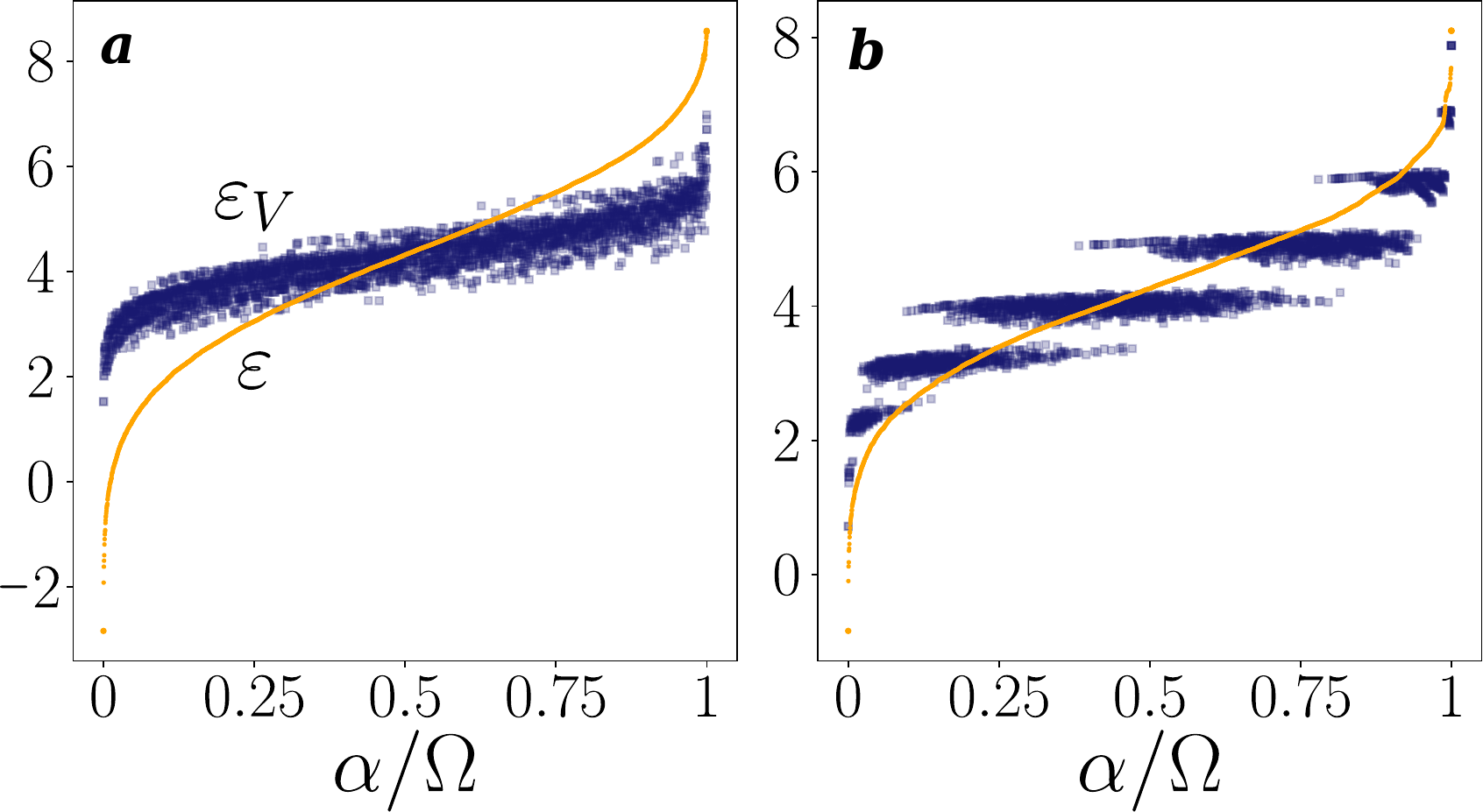} 
\centering
\caption{Emergence of plateaus. The scaled interaction energy $\varepsilon_V$ (blue symbols) for the eigenstates labeled by $\alpha=1, \cdots, \Omega$ sorted in ascending order of total energy $\varepsilon$ (orange symbols), where $\Omega$ is the dimension of the Hilbert space for $N/L=9/18$. (a) $\varepsilon_V$ and $\varepsilon$ for $V/t=2.2$; both reveal a continuum away from the borders. (b) $V/t=4.5$, where $\varepsilon_V$ forms discrete plateaus but $\varepsilon$ remains qualitatively the same.}
\label{fig_1}
\end{figure}

Let us first examine the energetic response of the eigenstates $|\varepsilon \rangle$ to increasing $|V|$. By exact diagonalization, we obtain $|\varepsilon \rangle$ and measure the scaled kinetic and interaction energies: $\varepsilon_K=E_K/V=-(t/V) \langle\varepsilon| \sum_j \hat{a}_{j+1}^{\dagger}\hat{a}_j+H.c. |\varepsilon \rangle$ and $\varepsilon_V=E_V/V=\langle \varepsilon |\hat{l}| \varepsilon \rangle$. The total energy is thus $\varepsilon=\varepsilon_K+\varepsilon_V$.
Without loss of generality, we always choose $V>0$ \cite{note_signV}.
The first observation is a distinct collective behavior of eigenstates as $V$ increases. Away from the edges of the spectrum, $\varepsilon$ sustains its continuous nature \cite{note_gap}, while $\varepsilon_V$ transitions from a continuum (Fig. \ref{fig_1}\textbf{a}) to a structured pattern of plateaus (Fig. \ref{fig_1}\textbf{b}). Each plateau is associated with an integer in the range $0, 1, \cdots, N-1$ approximating $\varepsilon_V$, meaning that the states constructing a plateau are spanned mostly by bases with the same number of NN links.

\begin{figure}[t]
\includegraphics[width=\columnwidth]{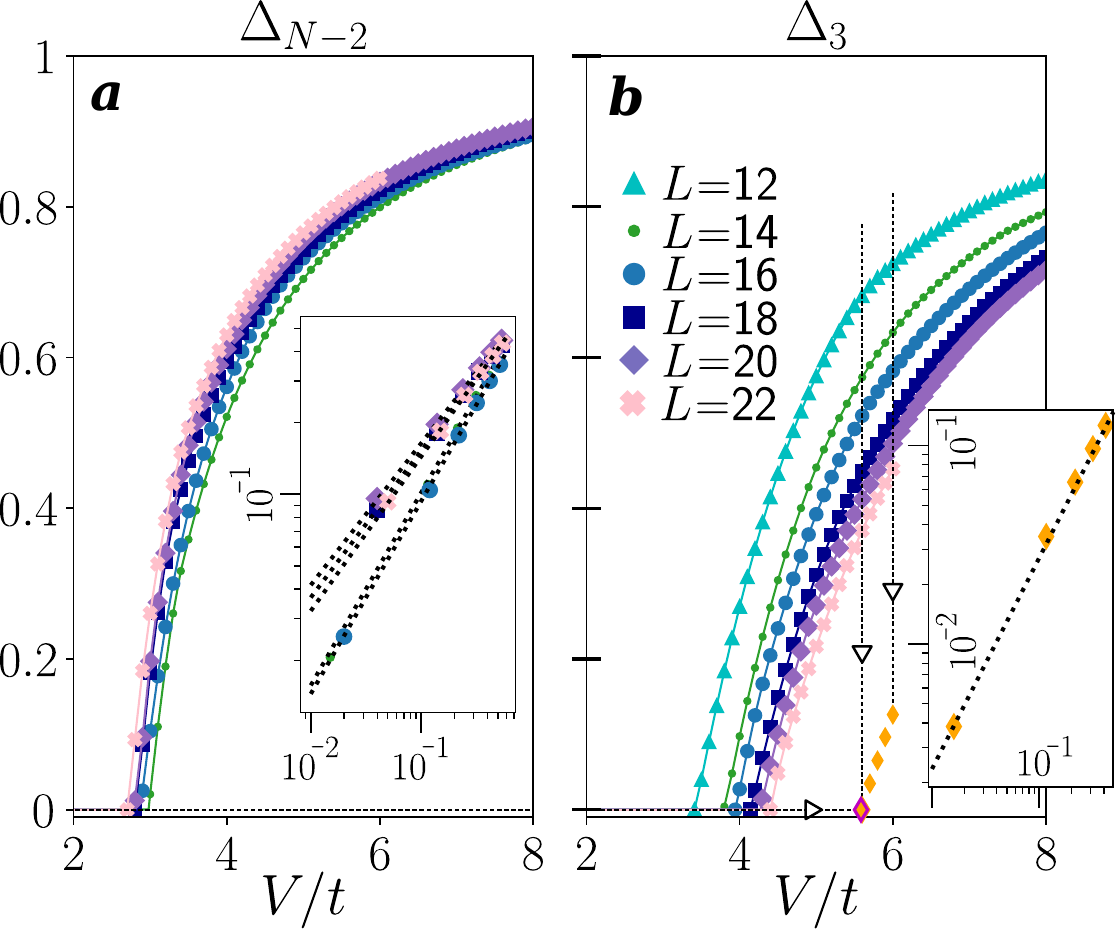} \centering
\caption{Plateau gaps and extrapolation in the fully symmetric subspace (see the text). (a) $\Delta_{N-2}$ versus $V/t$ for various system sizes. (b) $\Delta_{3}$; the arrows represent the linear extrapolation to $L\to\infty$. Insets: $\Delta_i$ versus $(V-V_c)/t$ on logarithmic scales; dashed lines confirm the power law relationship.}
\label{fig_2}
\end{figure}

An immediate question is how the gaps in $\varepsilon_V$ open as $V$ increases. Restricting our analysis to the fully symmetric subspace (zero quasi-momentum, even parity, and reflection symmetry), we estimate the $i{\textrm{th}}$ gap $\Delta_i$ (with $i=1$ to $N-1$, moving from left to right in Fig. \ref{fig_1}\textbf{b}) between two adjacent plateaus. Figure \ref{fig_2} illustrates examples of $\Delta_{N-2}$ and $\Delta_{3}$ as a function of $V/t$. For $\Delta_i$ values that decrease with $L$ (Fig. \ref{fig_2}\textbf{b}), we apply linear extrapolation to account for finite-size effects. To validate the extrapolation results, we consider, for each $L$, the value $V_0$ where the gap approaches zero. Extrapolation using $V_0$ yields an almost perfect self-consistent critical value $V_c$. For details, see Appendix A.

Curiously, the gaps behave similarly to order parameters in equilibrium critical phenomena presenting in continuous phase transitions: With a critical interaction strength $V_c$, the gap $\Delta_i$ is zero in the subcritical regime ($V<V_c$) and continuously increases to a nonzero value in the supercritical regime ($V>V_c$). Drawing an analogy with phase transitions, we examine the existence of universal power laws $\Delta_i \propto (V-V_c)^{\beta_i}$ as $(V-V_c) \to 0^{+}$, where $\beta_i$ is the critical exponent for the $i{\textrm{th}}$ gap. The insets in Fig. \ref{fig_2} show great agreement with the power-law behavior with impressive precision. 
Exact values for $V_c$ and $\beta_i$ should be interpreted with caution due to the limited system sizes. While the values of $V_c$ and $\beta_i$ appear to vary with $\Delta_i$, indicating distinct universal patterns associated with the eigenstates during plateau formation, whether this distinct universality will persist as $L \to \infty$ and $\Delta_{i \to \infty}$ has yet to be determined.

\section{Theoretical framework}
Despite the level-specific gaps, an arbitrary eigenstate can still preferentially occupy the subspace of a plateau. This concentration of eigenstates leads to the aggregation of $\varepsilon_V$ and supports the structure of plateaus across the spectrum. Based on perturbation analysis, we argue that the eigenstate concentration is stable in the thermodynamic limit. 
Moreover, the statistical properties of eigenstates give a more intuitive, although coarse-grained, insight compared to the gaps. We demonstrate that the extreme-value theory governs the statistics.

\subsection{Eigenvectors and preserving processes } \label{eigenvector}

For $V>t$ in the EHM, we consider $H/V = \hat{l} - (t/V)\, \hat{K}$, where $\hat{K} = \sum_{j} (\hat{a}^{\dagger}_{j+1}\hat{a}_j+\hat{a}^{\dagger}_{j}\hat{a}_{j+1})$ includes all the hopping, and $\hat{l}$ given by Eq. (\ref{l}) is diagonalized by Fock bases $|f\rangle$. 
Since $\langle f|\hat{l}|f \rangle = l_0 = 0,1,\ldots, N-1$ degenerates extensively, we combine $\hat{l}$ with the part of the hopping terms conserving the NN links. 
The model Hamiltonian then reads
\begin{equation}
\label{tv_H2}
    H/V = H_0 - (t/V)\widetilde{K}.
\end{equation}
The zeroth-order term $H_0$ governs the transition between Fock bases with the same $l_0$---, representing the singlon motions discussed above, while $\widetilde{K}$ transfers states with $l_0$ and $l_0 \pm1$.

Let $\{l_0\}$ be the subspace spanned by $|f\rangle$ with $\langle f|\hat{l}|f \rangle = l_0$; then $H_0=\widetilde{H}_{l=0}\bigoplus\widetilde{H}_{l=1}\bigoplus\cdots\widetilde{H}_{l=l_0}\cdots \widetilde{H}_{l=N-1}$. 
The sub-Hamiltonian $\widetilde{H}_{l_0}$ has constant diagonal terms $l_0$ and off-diagonal terms $-(t/V)$ representing the singlon motion. 
Recalling Eq. (\ref{N}), it is clear that if $\langle f |\hat{l}| f \rangle = 2,3\ldots, N-2$, $\langle f |\hat{c}| f \rangle$ and $\langle f |\hat{s}| f \rangle$ are not unique, implying sub-blocks within $\widetilde{H}_{l_0}$, but it does not affect the following analysis. 

\begin{figure}[t]
\includegraphics[width=\columnwidth]{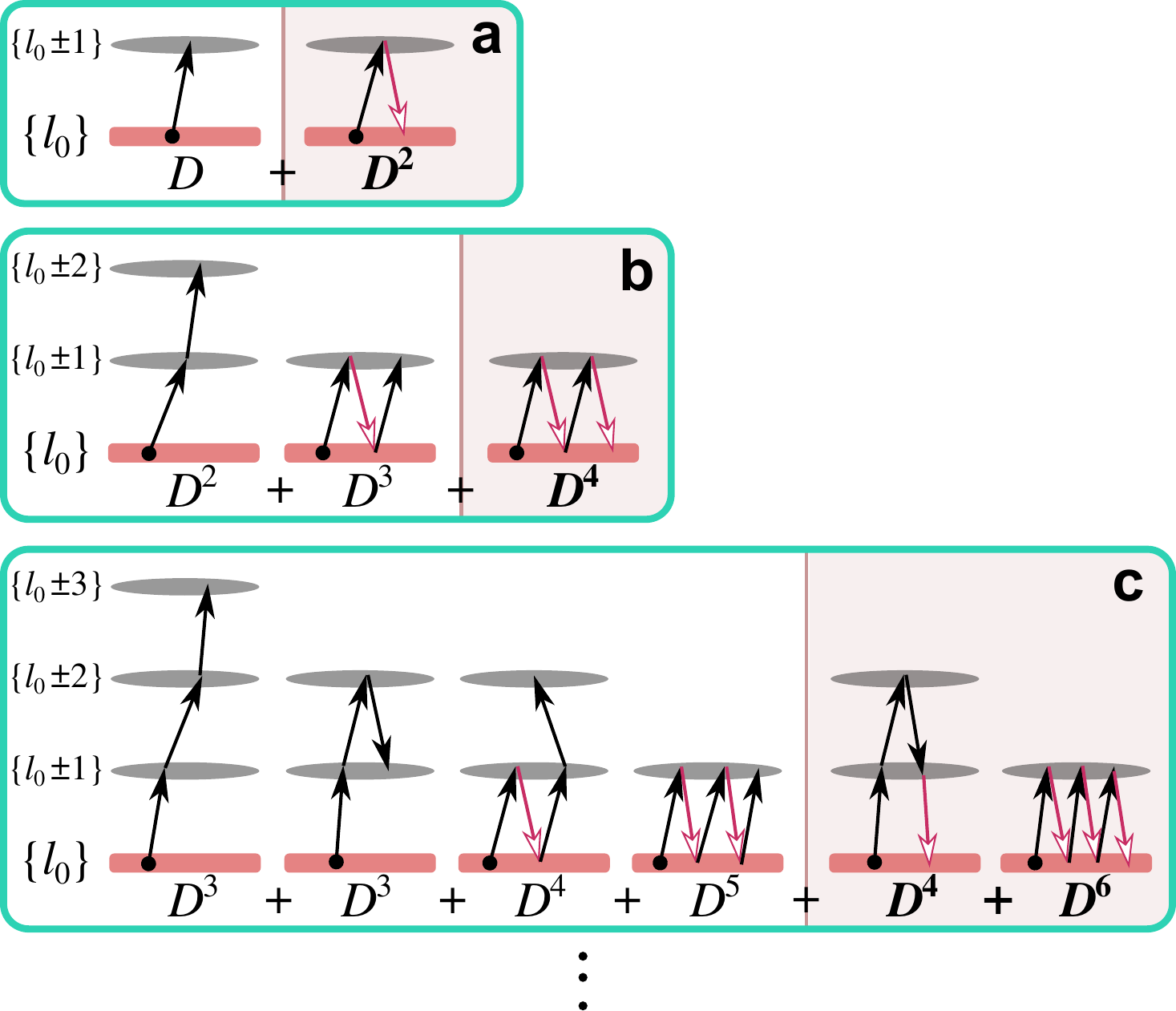} \centering
	\caption{Perturbation processes originate from $\widetilde{K}$. A process is a sequence of actions of $\widetilde{K}$, through which the zeroth-order state (black dots) is moved away from its subspace $\{l_0\}$ and transferred between adjacent subspaces. For a given process, the order of $(t/V)$ is the number of transfers  (arrows) except for those transferring from $\{l_0\pm 1\}$ back to $\{l_0\}$ (red arrows). The processes preserving $l_0$ are shown in the shaded areas. Powers of $D\propto L^2$ estimate the number of pathways for each process, a polynomial of $D$ is thus left for each order of $(t/V)$. (\textbf{a}), (\textbf{b}) and (\textbf{c}) show processes on the order of $(t/V)^m$ with $m=1$, $2$, and $3$, respectively.}
	\label{fig_S4}
\end{figure}

We expand the eigenstates as 
\begin{equation}
\label{expansion}
    | \varepsilon \rangle = |\varepsilon_0\rangle+\sum_{k=1}^{\infty}(t/V)^k\; | \varepsilon_k \rangle, 
\end{equation} 
where $|\varepsilon_0\rangle$ resides in $ \{l_0\}$ and is the zeroth-order state corresponding to the energy $\varepsilon_0$ given in Appendix B and $|\varepsilon_k\rangle$ is the term involving virtual processes that consist of $k$ actions of $\widetilde{K}$. As illustrated in Fig. \ref{fig_S4}, these actions move $|\varepsilon_0\rangle$ away from $\{l_0\}$ via transfers between adjacent subspaces.

The $k=1$ term is
\begin{equation*}
    |\varepsilon_1 \rangle 
    = \sum_{|\varepsilon'_0\rangle \in \{l'_0\} }
    \frac{ 
    \langle \varepsilon'_0 | \widetilde{K} | \varepsilon_0 \rangle
    }{\varepsilon_0-\varepsilon'_0} 
    \,
    |\varepsilon'_0 \rangle 
\end{equation*}
with $\{l'_0\}=\{l_0\pm1\}$. 
As illustrated in Fig. \ref{fig_S4}\textbf{a} (left panel), $|\varepsilon_1\rangle$ estimates those components of $|\varepsilon\rangle$ possessing $l_0 \pm 1$ NN links; it gives the leading contribution to the probability outside $\{l_0\}$.
The next term reads
\begin{equation*}
    |\varepsilon_2 \rangle 
    = 
    \sum_{ \substack{ |\varepsilon''_0\rangle \in \{l''_0\}\\ \varepsilon''_0 \;\neq\; \varepsilon_0 } }
    \hspace{0.15cm}
    \sum_{|\varepsilon'_0\rangle \in \{l'_0\} }
    \frac{
    \langle \varepsilon''_0 | \widetilde{K} | \varepsilon'_0 \rangle
    \,
    \langle \varepsilon'_0 | \widetilde{K} | \varepsilon_0 \rangle
    }
    {(\varepsilon_0-\varepsilon''_0) \, (\varepsilon_0-\varepsilon'_0)}
    \,
    |\varepsilon''_0 \rangle.
\end{equation*}
It involves two actions of $\widetilde{K}$, bringing $|\varepsilon_0\rangle$ intermediately to $| \varepsilon'_0 \rangle\in\{l'_0\}=\{l_0\pm1\}$ and then to $|\varepsilon''_0\rangle \in \{l''_0\}=\{l'_0\pm1\}$. 
When $\{l''_0\} = \{l_0\}$, the state $|\varepsilon_0\rangle$ is brought back to $\{l_0\}$ (Fig. \ref{fig_S4}\textbf{a}, right panel). In such a crucial situation, $(t/V)^2|\varepsilon_2\rangle$ is on the order of $(t/V)$ rather than $(t/V)^2$, as detailed in Appendix B; the denominator $(\varepsilon^{0}-\varepsilon''^{0})$ reduces by a factor of $(t/V)$ when both $\varepsilon^{0}$ 
and $\varepsilon''^{0}$ are eigenvalues of the same $\widetilde{H}_{l_0}$. 
This process preserves the number of NN links while enabling the number of singlons to change.\\

Further expansions are constructed systematically \cite{book_perturbation1, book_perturbation2}; Figs. \ref{fig_S4}\textbf{b} and \ref{fig_S4}\textbf{c} illustrate some of them. The processes preserving $l_0$ are highlighted in the right panels (shaded areas), where the last transfers (red arrows) always bring the state from $\{l_0\pm1\}$ back to $\{l_0\}$ and one factor of $(t/V)$ is always reduced by the involved denominator---, a difference between two eigenvalues for the same $\widetilde{H}_{l_0}$. As a result, the term $(t/V)^k|\varepsilon_k\rangle$ in Eq. (\ref{expansion}) is actually on the order of $(t/V)^m$, where $m=k-k'$ if $|\varepsilon_k\rangle$ contains $k'$ instances of $\{l_0\pm 1\}\to\{l_0\}$ transfers.

\subsection{ 
Plateau structures at the limit \texorpdfstring{$ \boldsymbol{L\to\infty} $}{thermo limit} 
}
\label{stability}
If the processes that return $|\varepsilon_0\rangle$ to $\{l_0\}$ outweigh those that move $|\varepsilon_0\rangle$ away from $\{l_0\}$, then 
the eigenstate will concentrate in $\{ l_0 \}$, and the interaction energy $\varepsilon_V$ will aggregate around $l_0$. 
This concentration is supported by two factors: First, maintaining $l_0$ of a zeroth-order state requires that higher-order corrections end up within $\{l_0\}$. The number of states in $\{l_0\}$ increases exponentially as $L \to \infty$, except at the borders where $l_0 = 0$ or $N-1$. Second, due to the degeneracy of $l_0$, for a given order of $(t/V)$, the processes that return a state to $\{l_0\}$ will predominantly contribute to the full eigenstates. The following is our argument. 

The leading violation of $l_0$ originates from $(t/V)|\varepsilon_1\rangle\sim\mathcal{O}(t/V)$ (Fig. \ref{fig_S4}\textbf{a}, left panel), which introduces a factor $D$, quantifying the number of pathways through $\widetilde{K}$ that create or annihilate a NN link in $|\varepsilon_0 \rangle$. 
According to Eq. (\ref{N}), a change in the number of NN links can come only from a change in singlons or from a change in clusters; hence, $D \propto \langle \varepsilon_0 | \hat{s}\hat{c} |\varepsilon_0\rangle $. Deep in the spectrum, both the number of singlons and the number of clusters scale with $L$, so $D \sim \mathcal{O}(L^2)$. 
The divergence of $D$ at the limit $L\to\infty$, however, does not negate the significance of $l_0$ due to those processes in $|\varepsilon_2\rangle$ that preserve $l_0$. 
As just discussed above, these processes reduce one $(t/V)$ according to the final states $|\varepsilon''_0\rangle \in \{l_0\}$ and introduce an additional factor $D' \propto \langle \varepsilon'_0 | \hat{s}\hat{c} |\varepsilon'_0\rangle$ via the intermediate states $|\varepsilon'\rangle \in \{l_0\pm 1\}$. Hence, $(t/V)^2|\varepsilon_2\rangle \sim \mathcal{O}(t/V)$, and an overall quantity of the pathways $D D' \sim D^2$.  

As illustrated in Figs. \ref{fig_S4}\textbf{b} and \ref{fig_S4}\textbf{c}, 
for a given order of $(t/V)^m$, the quantity of pathways for all processes is approached by $(D^{2m} + a_{2m-2}D^{2m-2} \cdots)+(a_{2m-1}D^{2m-1} \cdots + a_{m}D^m)$, where the first and second polynomials refer to the processes preserving and violating $l_0$, respectively. 
The highest-degree term, $D^{2m}$, contributes exclusively to preserving processes, which move $|\varepsilon_0\rangle$ back and forth between $\{l_0\}$ and $\{l_0\pm 1\}$ over $m$ times. We therefore argue that the preserving processes make an overwhelming contribution to the full eigenstates at $L \to \infty$. 
As a result, $\varepsilon_V$ aggregates around $l_0$ and a plateau emerges. 

\subsection{Extreme-value statistics}\label{GEV}
The quantum nature of the model allows nonzero wavefunction leaking outside $\{l_0\}$, resulting in a nonequivalence between $\varepsilon_V$ and $l_0$ even at $V/t \to \infty$, thereby widening the plateaus (Fig. \ref{fig_1}\textbf{b}). 
Through the statistics of these leakages, we provide a comprehensive description of the plateau structure.

\begin{figure}[t]	
\includegraphics[width=0.9\columnwidth]{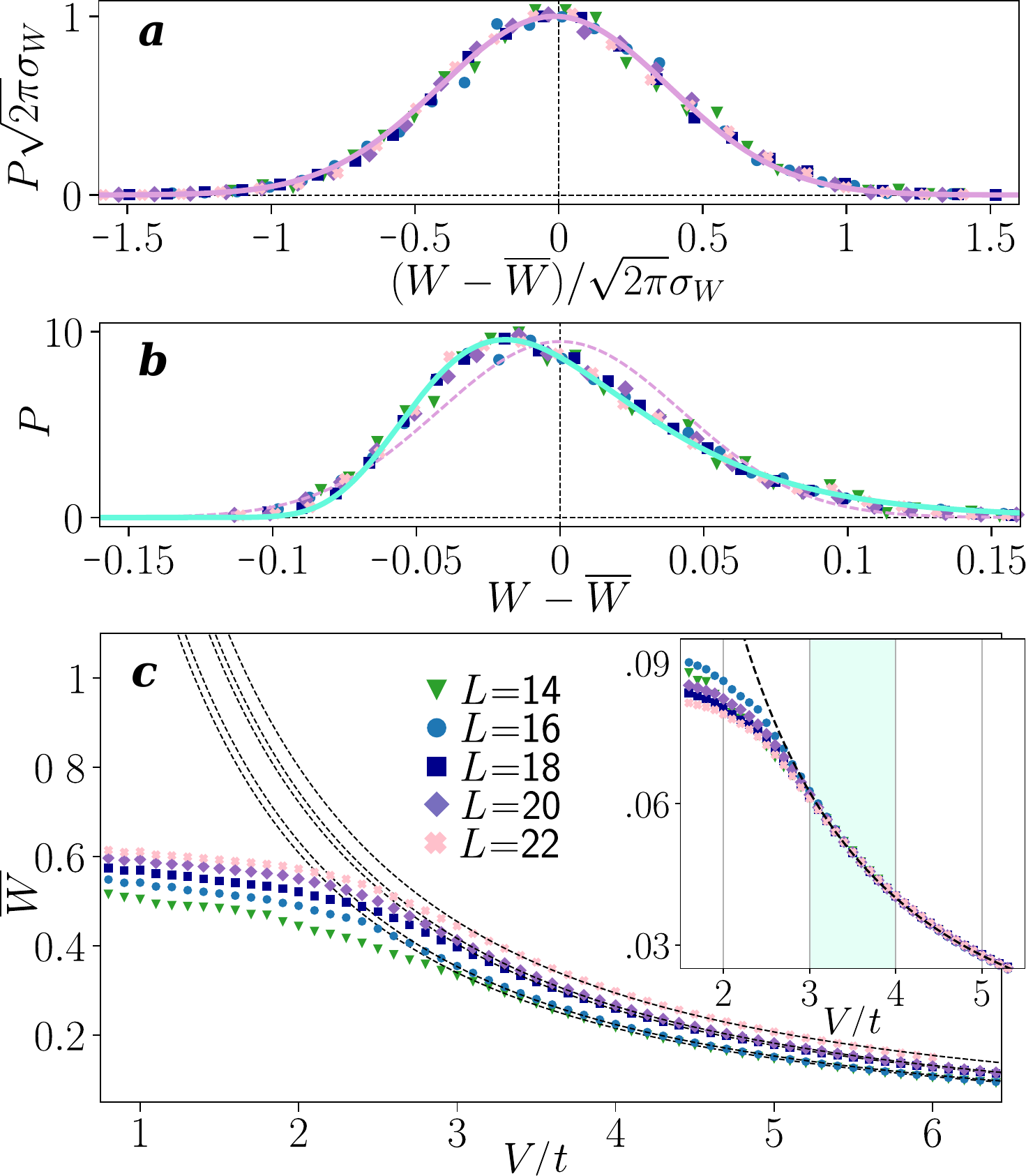} \centering
\caption{Emergent extreme value statistics.
 The results are based on the weight $W$ for eigenstates in the middle of the spectrum. Panels (a) and (b) show the probability density $P(W)$ for $V/t \sim 1.8$ and $V/t \sim 5.6$, respectively, obtained using the bin average method, where $\overline{W}$ represents the mean value and $\sigma_{W}$ denotes the standard deviation. Solid lines show the best fit to the data, where the Gumbel distribution is $P(x) = \frac{1}{\sigma_{W}}\exp[ -(x + e^{-x})]$ with $x=(W-\overline{W})/\sigma_{W}$ and $\sigma_{W}\sim 0.038(2)$. The dashed line in (b) is a normal-distribution fit for comparison; the displacement between the peaks of the two fittings is $\sim0.58(3) \sigma_W$. (c) $\overline{W}$ versus $V/t$; a decrease proportional to $ (V/t)^{-2}$ appears when $V/t$ exceed the threshold ($\sim 3.3$). The inset shows data for different $L$ collapses after a slight vertical shift.}
\label{fig_3}
\end{figure}

The underlying probability theory in our analysis here is validated since the level statistics of the XXZ model exhibit the Poisson distribution \cite{Poilblanc1993}. 
This indicates that the energy eigenvalues are characterized by independent random variables; we therefore also expect $P_{\varepsilon}(f) = |\langle f | \varepsilon \rangle|^2$, the weights of eigenstates $|\varepsilon\rangle$ on basis states $|f\rangle$, to reveal random behavior \cite{note_random}.
We compute the leakage 
\begin{equation}\label{leakage}
    W(\varepsilon)=\sum_{f'}|\langle f'| \varepsilon \rangle|^2,
\end{equation}
with bases $|f'\rangle$ satisfying $\langle f'|\hat{l}|f'\rangle \neq l_0$. 
Recalling the discussion in Sec. \ref{eigenvector}, $W$ is approximated by $(t/V)^2 |\langle \varepsilon_1| \varepsilon_1 \rangle|^2$, leading to the expectation that $W \propto (V/t)^{-2}$. Fig. \ref{fig_3}\textbf{c} shows the mean value $\overline{W} = \frac{1}{\Omega} \sum_{\varepsilon} W(\varepsilon)$ as a function of $V/t$. The observed relationship $W \propto (V/t)^{-2}$ starts at $V/t \gtrsim 3.5$; in agreement with our estimation of the threshold values, see Appendix C for details.

Figs. \ref{fig_3}\textbf{a} and \ref{fig_3}\textbf{b} illustrate that the probability distribution $P(W)$ transitions from a normal form to a distorted one after $V/t$ surpasses the threshold. 
For an eigenstate described by the weight $P_{\varepsilon}(f) = |\langle f | \varepsilon \rangle|^2$, the leakage $W$ simulates the mean of the samples: $P_{\varepsilon}(f=f')$.
A weaker interaction allows eigenstates to spread naturally across the Hilbert space, with $P_{\varepsilon}(f')$ thus acting as a random sampling that gives arbitrary values in $P_{\varepsilon}(f)$. According to the central limit theorem, the statistics of the mean of the random samples, i.e., $P(W)$, lead to a normal distribution as shown by Fig. \ref{fig_3}\textbf{a}.
However, if eigenstates concentrate in $\{l_0\}$, the leakage decreases, and $P_{\varepsilon}(f')$ no longer presents random samples; instead, it goes to vanishing values in the course of the concentration. The statistics of $P_{\varepsilon}(f')$ in such a scenario lead to the statistics of minimums, described by the generalized extreme value theory \cite{Majumdar2020}. Recalling the discussion of the perturbation expansion, it is evident that the value of $P_{\varepsilon}(f')$ for $| f' \rangle \in \{l'_0 \neq l_0 \}$ scales as $(V/t)^{-|l'_0 - l_0|}$, which decays exponentially as $l'0$ moves away from $l_0$. Consequently, for a significant portion of eigenstates $| \varepsilon \rangle$, particularly those in the middle of the spectrum, $P{\varepsilon}(f')$ follows a distribution with an exponentially decaying tail. According to extreme-value theory \cite{Majumdar2020}, regardless of this distribution's specific form, the minimum values' statistics will conform to the Gumbel distribution. Fig. \ref{fig_3}\textbf{b} demonstrates good consistency between this distribution and our dataset.

\section{Universal dynamics}\label{dynamics}
In the presence of the plateau structure, the diminishing leakage of eigenstates from the subspace $\{l_0\}$ implies a reduction in the traversed area in the Hilbert space during time evolution; 
the dynamical feature is captured by the Fock-state configurations in $\{l_0\}$ determined by a fixed $N/L$. 
As demonstrated below, the evolution of Fock states conserves $\hat{l}$ to first order in $(t/V)$, and the local equilibration is approximated by a function of time that is independent of $V$ after appropriate normalization.

\subsection*{ Time evolution with quasiconserved \texorpdfstring{$ \boldsymbol{l_0} $}{l zero} }
The time evolution of an arbitrary Fock state $|f\rangle$ is given by $ \langle f' |f(\tau) \rangle=\sum_{\varepsilon} e^{-i \varepsilon \tau}  \langle f' | \varepsilon \rangle\langle \varepsilon| f \rangle $, where
\begin{equation}
\label{projector}
\begin{split}
| \varepsilon \rangle\langle \varepsilon| = \;
&| \varepsilon_0 \rangle \langle \varepsilon_0|\\
&+(t/V) \, \Bigl(|\varepsilon_0 \rangle \langle \varepsilon_1| + | \varepsilon_1 \rangle \langle \varepsilon_0| \Bigr)\\
&+ (t/V)^2 \, \Bigl(|\varepsilon_1 \rangle \langle \varepsilon_1|+|\varepsilon_0 \rangle \langle \varepsilon_2|+|\varepsilon_2 \rangle \langle \varepsilon_0| \Bigr) \cdots.
\end{split}
\end{equation}
Equation (\ref{expansion}) is used for the above expansion.
For the initial state $|f\rangle$, we consider $|f\rangle \in \{l_0\}$, thenwe have the following:

\begin{itemize}
\item [1.] If $|f'\rangle \in \{l_0\}$ then $\langle f' |f(\tau) \rangle \sim \mathcal{O}(t/V)$ since $|\varepsilon_0 \rangle\langle \varepsilon_0|$ yields zero.
\item [2.] If $|f'\rangle \in \{l_0\}$ then $\langle f' |f(\tau) \rangle \sim \mathcal{C}(\tau) + (t/V) \cdot \mathcal{F}(\tau)$, 
\end{itemize}
where 
$$\mathcal{C}(\tau) = \sum_{\varepsilon} e^{-i\varepsilon\tau} \langle f' |\varepsilon_0 \rangle \langle \varepsilon_0 | f \rangle $$ 
describes the singlon motions conserving both the number of NN links and the number of singlons, 
$$\mathcal{F}(\tau) = \sum_{\varepsilon} e^{-i\varepsilon\tau} (\langle f' |\varepsilon_0 \rangle \langle \tilde{\varepsilon}_2 | f \rangle+\langle f' |\tilde{\varepsilon}_2 \rangle \langle \varepsilon_0 | f \rangle)$$ 
describes the first-order dynamics that conserve only the NN links, and $|\tilde{\varepsilon}_2\rangle$ represents the part of $|\varepsilon_2 \rangle$ preserving $l_0$. 
This is because  $|\varepsilon_0 \rangle \langle \varepsilon_1|$ and $|\varepsilon_1 \rangle \langle \varepsilon_0|$ give zeros and  $|\varepsilon\rangle \langle\varepsilon| \sim |\varepsilon_0\rangle \langle\varepsilon_0| + (t/V) \,(|\varepsilon_0\rangle \langle\tilde{\varepsilon}_2|+|\tilde{\varepsilon}_2\rangle \langle\varepsilon_0|)$, see Sec. \ref{eigenvector}. 
\\

The terms $\mathcal{C}(\tau)$ and $\mathcal{F}(\tau)$ are approximately independent of $V$ because the energy eigenvalues $\varepsilon$ in the model behave as independent random variables (see Sec. \ref{GEV}). Consequently, the term $\sum_{\varepsilon} e^{-i\varepsilon \tau} \cdots$ is expected to be approximated by a random phase average; the features of $\mathcal{C}(\tau)$ and $\mathcal{F}(\tau)$ are thus determined by the Fock-state configurations in $\{l_0\}$, which is defined by a given $N/L$.
\\

Observables that are characterized by $|\langle f' | f(\tau) \rangle|^2$ thus exhibit the following behavior: if $|f\rangle$ and $|f'\rangle$ possess the same number of NN links, then the observable scales as $|\mathcal{C}|^2+\mathcal{O}(t/V)$; otherwise, it scales as $\mathcal{O}(t^2/V^2)$. Therefore, the dynamics are predominantly restricted by the initial number of NN links to the order of $(t/V)$, even as the singlon number may evolve.
\begin{figure}[t]
\includegraphics[width=\columnwidth]{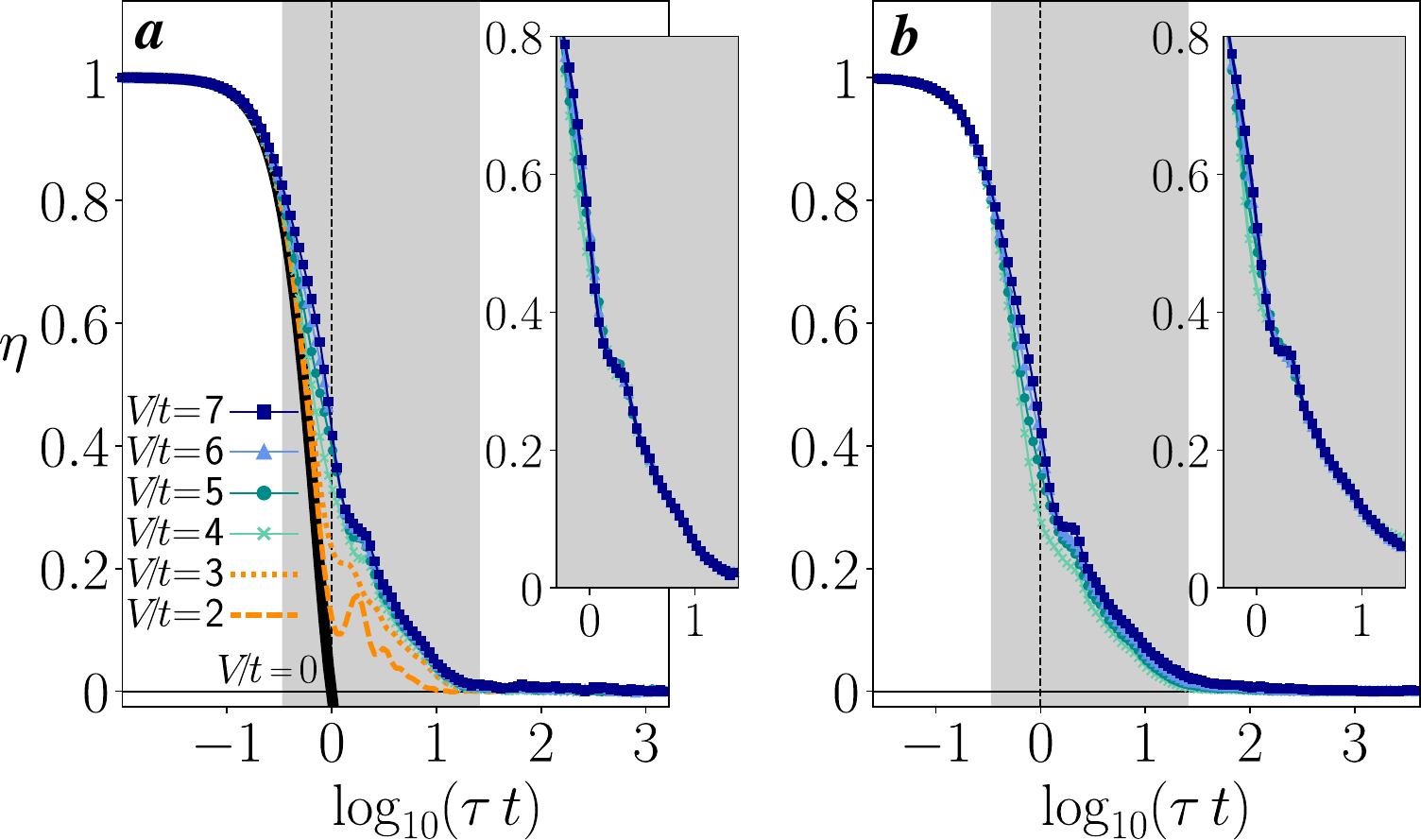} \centering
\caption{Equilibration of Fock states. (a) $\eta(\tau)$ for $L=18$; two stages of dynamics are observed before and after $\tau t \sim 1$. The inset shows the universal form $A (\tau)$ with $\epsilon \sim 0.05$ (see the text) governing the dynamics for $V/t \gtrsim 3.3$. (b) Similar to (a) but with dipole-dipole interactions $V_{i,j} = V/|i-j|^{3}$ and $\epsilon \sim 0.25$.}
\label{fig_4}
\end{figure}

\subsection*{Local equilibration}
To align with experiments, we investigate this impact by examining the equilibration of Fock states $|f_0\rangle$ with the $j_0{\textrm{th}}$ site initially occupied. We compute $\eta(\tau) = [\langle f_0(\tau)|\hat{n}_{j_0}| f_0(\tau) \rangle-N/L] / (1-N/L)$, where $\eta=1$ indicates that the initially observed particle is localized at $j_0$, while $\eta=0$ implies that it has dissipated into the background. Assuming that the initial preparation \cite{Bakr2009, Kuhr2016, Gross2021} does not favor particular configurations, we average $\eta(\tau)$ over all $|f_0 \rangle$ for a statistical description.
The dynamics depicted in Fig. \ref{fig_4}\textbf{a} unveil two stages. For $\tau t \lesssim 1$, free-moving singlons dominate the dynamics, yet the final equilibration at $\eta=0$ is not achieved, contrasting with the behavior of free-moving particles with $V=0$. This suggests that singlons alone do not furnish sufficient motion to connect $|f_0\rangle$ equally to $j_0$-occupied and $j_0$-empty states.
Nonetheless, equilibration persists. For $\tau t \gtrsim 1$, the spread of singlons saturates, uncovering the remaining higher-order processes. These processes compel equilibration to progress until $\eta=0$.

A slight variation of $\eta(\tau)$ is observed with different values of $V/t$ when it exceeds the threshold, reflecting the dependence of higher-order processes on the interaction strength. The first-order approximation allows us to express $\eta(\tau) \sim A(\tau) - B(\tau) \, (t/V)$, where $A(\tau)$ and $B(\tau)$ are $V$-independent functions related to dynamics conserving NN links. 
The expressions for $A(\tau)$ and $B(\tau)$ can be approached by expanding $\eta(\tau)$ to first order in $(t/V)$; 
note that $\eta(\tau) \propto \frac{1}{\Omega_0}\sum{f_0, f'_0} \left| \langle f'0| f_0(\tau)\rangle \right|^2 \sim \frac{1}{\Omega_0}\sum{f_0, f'_0} \left| \mathcal{C}(\tau) + (t/V) \mathcal{F}(\tau) \right|^2$. 
Given that both $A(\tau)$ and $B(\tau)$ stem from processes preserving $l_0$, which are determined by basis configurations in $\{l_0\}$, we take the simplification $B(\tau) \sim A(\tau)+\epsilon$. 
As a result, the dynamics can be described by the universal form $A(\tau) \sim [\eta(\tau) + (t/V)\epsilon] / [1 - (t/V)]$ (Fig. \ref{fig_4}\textbf{a}, inset). 

\section{Conclusions} 
We studied a generic 1D quantum lattice model and uncovered a profound relationship between out-of-equilibrium dynamics and emergent properties in eigenstates across the spectrum. Specifically, a strong but finite, intersite interaction induced inherent structures within the eigenstates, leading to the reorganization of wavefunctions across the Fock bases and the opening of gaps in interaction energy akin to continuous phase transitions. These structured eigenstates exhibit statistical patterns described by extreme-value theory and compress the area of time evolution, wherein the equilibration of Fock states unfolds through multiple stages with a universal nature.

Our findings, particularly regarding the dynamical aspects, demonstrate significant consistency across a broad range of realistic interactions (e.g., Fig. \ref{fig_4}\textbf{b}), suggesting that our observations are not a consequence of the nonthermal feature, i.e., the integrability, originating from the NN interacting form \cite{quantum_integrability}. This offers a comprehensive understanding of recent observations \cite{Sanchez2021}, wherein a qualitative slowdown of quench expansion was confirmed with interaction strengths close to those associated with plateau structures.

The EHM under examination is isomorphic to the XXZ model, where the structured eigenstates provide deeper insight into the transformation of the self-return probability \cite{Misguich2016} observed previously.
Spin transport in the XXZ model has been extensively studied \cite{Bertini2021, Landi2022}. Earlier research observed behavior consistent with linear response theory in far-from-equilibrium scenarios \cite{Steinigeweg2017, Ljubotina2017}, where transport is ballistic for $V/t < 2$, superdiffusive [Kardar-Parisi-Zhang (KPZ)] at $V/t = 2$, and diffusive for $V/t > 2$. However, this behavior depends on initial conditions that reflect specific equilibrium backgrounds in terms of decoherence and/or entanglement \cite{Steinigeweg2017}. In contrast, the dynamics of initial Fock states avoid these limitations, thus exploring aspects of far-from-equilibrium dynamics that are not accessible through linear response theory. The reduced evolving area of Fock states presents weakened ergodicity due to interactions, a phenomenon that should be distinguished from non-thermal states arising from kinetic constraints \cite{QMBS, localization}.


\textit{Acknowledgments.---} 
We thank R. Moessner for the insightful discussions and his constructive comments throughout the completion of this project.

\section*{APPENDIX}
Here we provide details for plateau structures including extrapolation results (Sec. \ref{app_plateau}), the analysis of zeroth-order energy for $\widetilde{H}_{l_0}$ (Sec. \ref{app_subH}), and the estimation of the threshold $V/t$ for the plateau structure (Sec. \ref{app_threshold}). 
\appendix
\section{Details for the plateaus}\label{app_plateau}
\begin{figure}[b]
\includegraphics[width=0.8\columnwidth]{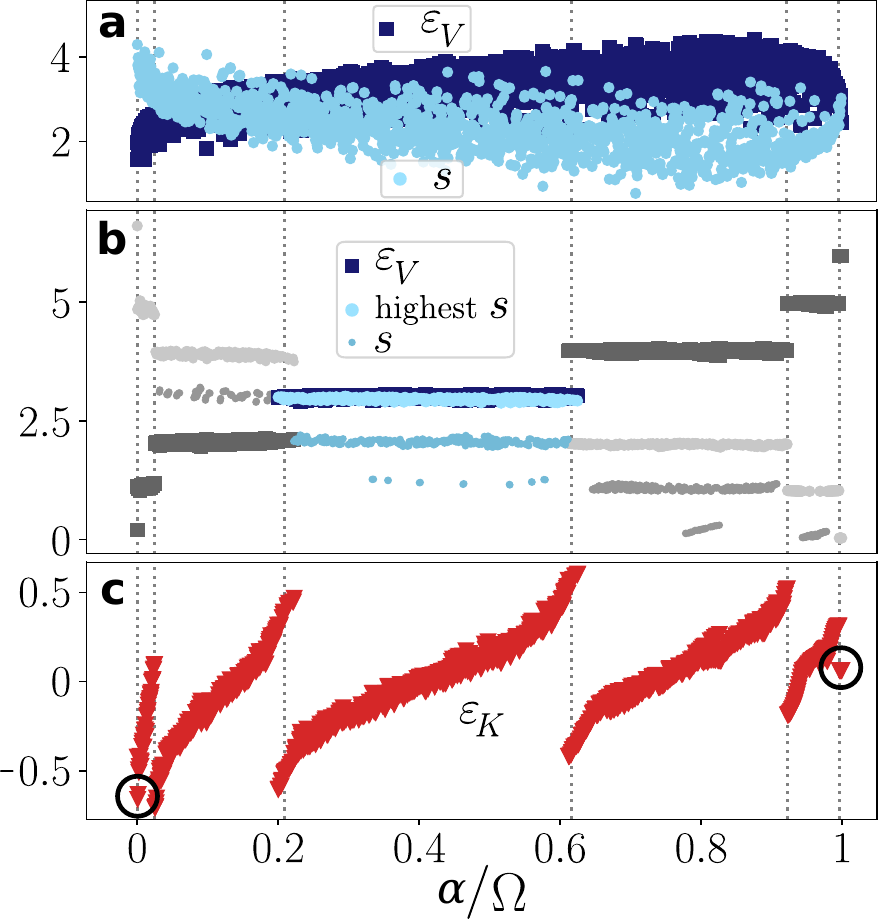} \centering
	\caption{Details of emergent plateaus. 
 The scaled interaction energy $\varepsilon_V$ (dark $\blacksquare$), expected singlon number $s$ (light $\bullet$), and scaled kinetic energy $\varepsilon_K$ (red $\blacktriangledown$) for the eigenstates labeled by $\alpha=1, \cdots, \Omega$ sorted in ascending order of total energy $\varepsilon = \varepsilon_V+\varepsilon_K$, where $\Omega$ is the dimension of the Hilbert space for $N/L=7/14$. 
(\textbf{a}) $\varepsilon_V$ and $s$ for $V/t=1$.
(\textbf{b}) $\varepsilon_V$ and $s$ for $V/t=8$; eigenstates exhibit a common structure when their $\varepsilon_V$ falls within the same plateau, and $s$ is at the same level. Note that $s$ is at its maximum level on the two sides of each plateau.
(\textbf{c}) $\varepsilon_K$ for $V/t=8$; bandwidths extend to the middle of the total energy spectrum but approach zero at the edges (circles). }
	\label{fig_6}
\end{figure}
As mentioned in the main text, an increasing $|V|$ suggests structures in eigenstates $|\varepsilon\rangle$ that can be mainly identified by $\varepsilon_V(\varepsilon)=\langle \varepsilon | \hat{l} | \varepsilon \rangle$; 
this is demonstrated in Figs. \ref{fig_6}\textbf{a} and \ref{fig_6}\textbf{b} by the plateaus of $\varepsilon_V$. Nonetheless, the formation of plateaus is accompanied by discrete levels of $s=\langle \varepsilon | \hat{s} | \varepsilon \rangle$, as basis states with the same number of NN links can have different numbers of singlons. Eigenstates belonging to the same plateau of $\varepsilon_V$ with the same level of $s$ can be considered to have a common structure; note that states belonging to different plateaus can never have the same structure even though they may have the same $s$. We take $V/t>N$  in Fig. \ref{fig_6}\textbf{b} for a better illustration; a weaker $V/t$ results in the same qualities.

Within a plateau, states at both edges consistently exhibit the highest singlon count (Fig. \ref{fig_6}\textbf{b}) and extreme values of kinetic energy $\varepsilon_K$ (Fig. \ref{fig_6}\textbf{c}). Since the interaction energies within a plateau closely approximate $l_0$, states on either side with the highest or lowest total energy exhibit extreme kinetic energies, indicating high-mobility states characterized by $|\varepsilon_K|$. At the boundaries of the total energy spectrum, $\varepsilon_V$ represents the total energy as the bandwidth of $\varepsilon_K$ vanishes. However, $\varepsilon_K$ diverges in the middle of the spectrum, resulting in a continuous spectrum as shown in the main text. We will revisit this topic in Appendix \ref{app_subH}.
\begin{figure}[t]
\includegraphics[width=\columnwidth]{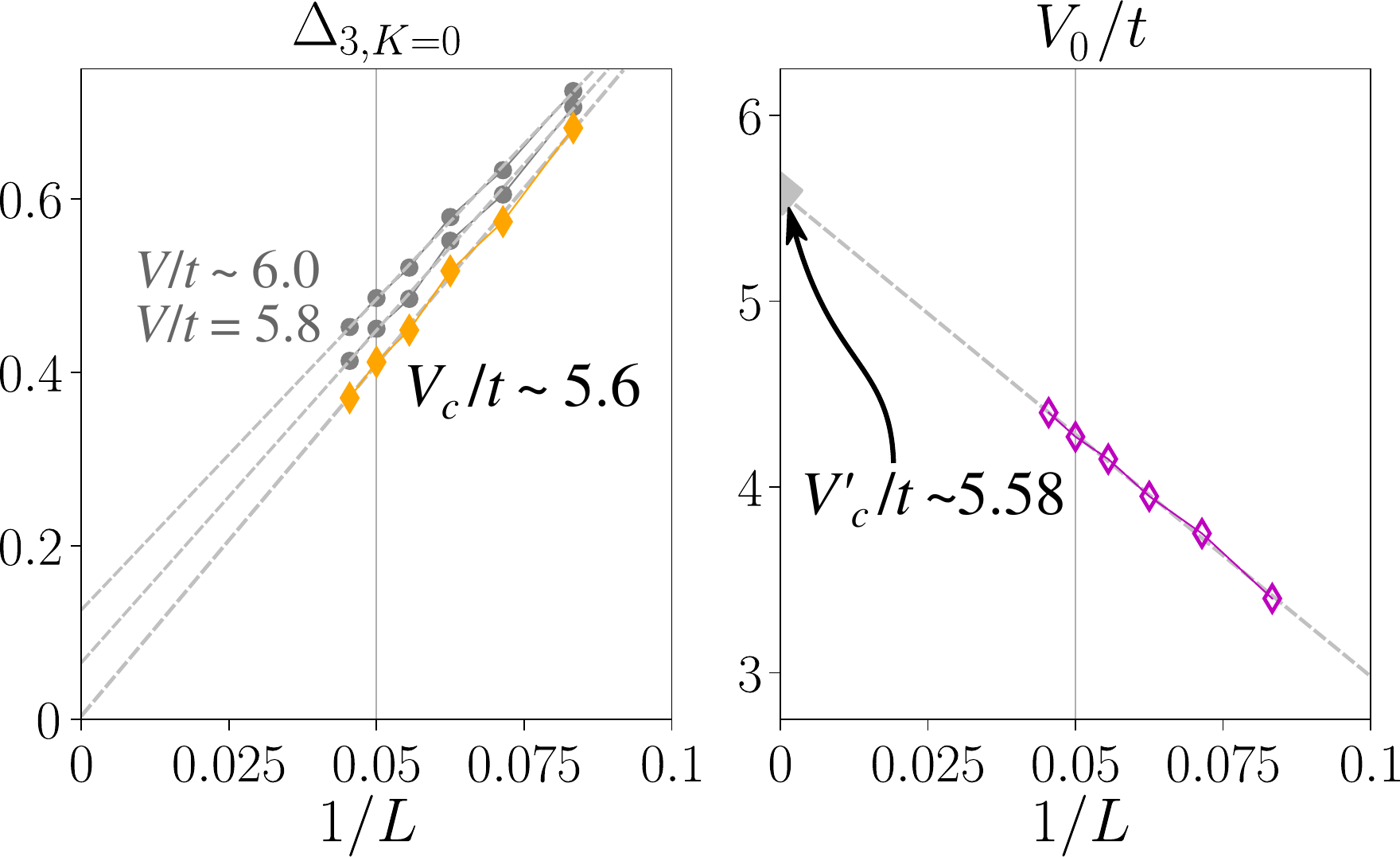} \centering
	\caption{The extrapolation. Concerning finite-size effects, the gap values are estimated for various $V/t$ by stretching $\Delta_i$ as a function of $1/L$. Data are shown for $\Delta_3$.}
	\label{fig_7}
\end{figure}

As the main text illustrated, the gaps behave similarly to order parameters of continuous phase transitions. 
To investigate the critical behavior of gaps $\Delta_i$ within the continuum of $\varepsilon_V$, we utilized the zero quasimomentum subspace ($K = 0$) with even parity and reflection symmetry, allowing us to perform exact diagonalization calculations up to $L=22$. To minimize finite-size effects, we conducted extrapolations for gaps near the center of the spectrum. For a self-consistent analysis, we focused on the gap $\Delta_i$ with $i\leq N/2$, specifically considering $\Delta_3$ for $L$ ranging from 12 to 22. Figure \ref{fig_7} (left panel) shows the linear extrapolation of the gap values as a function of $1/L$, with a critical value $V_c$ estimated. The dashed lines represent the best linear fits, and the standard deviation errors between the fits and the data are approximately $0.1\%$. 
To further validate the results obtained from the gap extrapolation, we analyzed the interaction $V_0$ for each $L$ where the gap approaches zero. The extrapolation using $V_0$, shown in Fig. \ref{fig_7} (right panel), results in a value $V'_c$ that aligns almost perfectly with $V_c$. Notably, the value obtained, $V_c/t \sim 5.6$, is consistent with the thermodynamic picture, where $V/t$ is predominantly determined by $L$ (and $N$).

\section{
\texorpdfstring{ $\boldsymbol{ \widetilde{H}_{l_0} } $ }{sub-H} 
and 
\texorpdfstring{ $\varepsilon_0$ }{e0}
}\label{app_subH}
The sub-Hamiltonian $\widetilde{H}_{l_0}$ has constant diagonal terms $l_0$ and off-diagonal terms $-(t/V)$ describing the amplitude of singlon hopping. 
The singlon motions conserve $l_0$; in other words, they are not allowed to be adjacent, either to each other or to other particles. This means singlons live in reduced lattices,
\begin{equation}
\label{Ls}
    L_s \sim L-N+l_0.
\end{equation}
Here we let $N_s=\langle f|\hat{l}|f \rangle$, with $|f\rangle \in \{l_0\}$. $\widetilde{H}_{l_0}$ is equivalent to the EHM in the main text with $N_s$ particles, $L_s$ sites, and $V=0$, which can be mapped into the XX model with magnetization $L_s-2N_s$ and zero transverse fields. The eigenvalues are given \cite{book_int_model} as 
\begin{equation}
\label{e0}
\varepsilon_{0,a} = l_0 + (t/V) \;\; \Lambda_a, \, \Lambda_a = -2\sum_{j=1}^{N_s} \cos(\frac{2\pi m_j}{L_s}),
\end{equation} 
where $l_0=\varepsilon_{V0}$ and $(t/V)\Lambda_a=\varepsilon_{K0}$ are the zeroth orders of $\varepsilon_V$ and $\varepsilon_K$ respectively. Note that as the diagonal terms of $\widetilde{H}_{l_0}$ are constants, $\Lambda_a$ can also be given by applying the Jordan-Wigner transformation to $\widetilde{H}_{l_0}-l_0$, followed by a Fourier transformation.

The lower index $a$ labels a total of $\binom{L_s}{N_s}$ states in $\widetilde{H}_{l_0}$, and the number $m_j$ indicates nonrepeating integers selected from the range $[0, L-1]$ for fermions or an odd number of bosons, but half-integers selected from $[\frac{1}{2}, L-\frac{1}{2}]$ for an even number of bosons. 
Choosing $m_j = 0,\ldots, N_s-1$ for the former case gives the lowest energy of Eq. (\ref{e0}),
\begin{equation}
\label{Lambda_min}
    -\Lambda_{min} 
    = 2\cos\left[\frac{\pi(N_s-1)}{L_s}\right] \frac{\sin(\pi N_s/L_s)}{\sin(\pi/L_s)}. 
\end{equation} 
Note that choosing $m_j = \frac{1}{2},\ldots, N_s-\frac{1}{2}$ for the latter case leads to the same result at $L_s\to\infty$. 
The bandwidth for eigenstates of $\widetilde{H}_{l_0}$ is $\Delta \varepsilon_0 =\Delta \varepsilon_{K0}= (t/V)(\Lambda_{max}- \Lambda_{min}) = (2t/V)|\Lambda_{min}|$. 

Within a plateau, $\varepsilon_V $ approaches $l_0$; thus, $ \Delta \varepsilon $ is close to $\Delta \varepsilon_K \sim \Delta \varepsilon_{K0}\propto |\Lambda_{min}|$.
At the left and right edges of the total energy spectrum, the Fock bases involved are of the types $|\bullet\circ\bullet\circ\cdots\bullet\circ \rangle$ and $|\bullet\bullet\cdots\bullet\circ\circ \cdots \circ \rangle$, respectively. For these configurations, $N_s=L_s$ or $0$, leading to $\Delta \varepsilon_K \to 0$. In contrast, in the middle of the spectrum, $\Delta \varepsilon_K$ scales as $\frac{1}{\sin(\pi/L_s)}$, which diverges in the thermodynamic limit. This divergence is numerically confirmed in Fig. \ref{fig_6}\textbf{c}. As a result, the total energy $\varepsilon = \varepsilon_V + \varepsilon_K$ forms a continuous spectrum, despite the presence of gaps in $\varepsilon_V$, as illustrated in Fig. 1 of the main text.

\section{Threshold estimation}\label{app_threshold}
The aggregation, $\varepsilon_V \sim l_0 = 0,1\ldots, N-1$, tells us that the interaction energy dominates the pattern of the total energy spectrum; precisely, $\varepsilon_V$ is significantly larger than $\Delta \varepsilon_K$. To estimate the required $V/t$, we start with 
\begin{equation}
\label{condition_0}
l_0 > \Delta \varepsilon_{K0},
\end{equation}
where $\Delta \varepsilon_{K0}$ is determined by Eq. (\ref{Lambda_min}). 
Since $\Delta \varepsilon_{K0} \to 0$ at the edges of the spectrum, we shall focus on the middle of it, where $\Delta \varepsilon_{K0}$ diverges as $\frac{1}{\sin{(\pi/L_s)}}$. 

For $l_0$ to significantly exceed $\Delta \varepsilon_{K0}$, 
we demand $l_0 > (4t/V)\,\left|\frac{\sin(\pi N_s/L_s)}{\sin(\pi/L_s)}\right|$, which serves as a sufficient condition of the inequality (\ref{condition_0}) capturing the divergence of $\Delta \varepsilon_{K0}$ in the middle of the spectrum. 
The scenario $l_0=N/2$ dominates the half-filling lattices at the thermodynamic limit since the size of $\{l_0\}$ decreases exponentially when $l_0$ departs from $N/2$. 
Let $l_0 = N/2$, which is followed by $L_s =L-N+l_0= 3N/2$; according to Eq. (\ref{Ls}), we have 
\begin{equation*}
V/t > \frac{4}{N/2} \left|\frac{\sin(2\pi N_s/3N)}{\sin(2\pi/3N)}\right|.
\end{equation*}
The bandwidths, determined by the extreme kinetic energies, naturally imply the largest number of singlons allowed in the plateau (as confirmed in Fig. \ref{fig_6}\textbf{c}). The value of $N_s$ is thus obtained by Eq. (\ref{N}) with $\hat{c}=1$, which gives $N_s = N-l_0-1$.
And since $l_0=N/2$, the above inequality results in $V/t \gtrsim 3.31$ after taking the limit $N \to \infty$. The estimation agrees well with the results in the main text.





\begin{thebibliography}{99}
\bibitem{note_mapping} 
Eq. (\ref{tv_H}) is also called the $t$-$V$ model; it describes a hard-core Bose lattice gas, which can be mapped into the XXZ model by Holstein–Primakoff transformation, so half filling is associated with zero magnetization. In one dimension, the XXZ model can again be mapped into a spinless Fermi lattice gas by Jordan-Wigner transformation.
\bibitem{note_signV} For the XXZ model \cite{note_mapping}, consider $\mathcal{U}=\exp \left( { i\pi \sum_{j} j \hat{S}_{j}^{z}}\right)$, a rotation by $\pi$ about the $z$ axis for every second spin, where $\hat{S}^z = \frac{1}{2}\hat{\sigma}^z$. The unitary transform $\mathcal{U}^{\dagger}H(V)\mathcal{U} = -H(-V)$ gives the spectrum relationship $E(-V) = -E(V)$. So $E(-V)$ and $E(V)$ are exact reflections of each other to $E = 0$, note that the ground state for $-V$ corresponds to the highest excited state for $V$. Hence, to investigate $E(V)$ with a varying $|V|$, taking $V>0$ will be enough. Although the sign of $V$ may result in different details in the eigenstates, the time evolution given in Sec. \ref{dynamics} is qualitatively the same, especially after averaging over different initial configurations.
\bibitem{note_gap} For $|V/t|>2$, it is well known that the ground state is gapped \cite{quantum_integrability}, but close to its center, the spectrum is gapless at the limit $L\to\infty$. See Appendix B for an estimation.
\bibitem{note_random} In practical terms, this means that the distribution of these weights should reflect the randomness seen in the eigenvalue statistics. For instance, if the energy levels of the XXZ model follow a Poisson distribution, the probability of finding an eigenvector predominantly in any given basis state should also be distributed randomly. This aligns with the notion that eigenvectors are spread out across the basis states without a discernible pattern, mirroring the randomness observed in the level statistics.
\end{thebibliography}
\end{document}